\documentclass[twocolumn,preprintnumbers,prd]{revtex4}
\usepackage{amssymb}
\usepackage{amsfonts}
\usepackage{amsmath}
\usepackage{graphicx}
\usepackage{dcolumn}
\usepackage{bm}
\usepackage{acronym}
\usepackage{caption2}

\begin{document}

\title{MHD waves within noncommutative Maxwell theory}
\author{S. Bourouaine\footnote{On leave of absence from Department of
 Physics, Faculty of Science, Mentouri University, Constantine,
 Algeria. Email: bourouaine@mps.mpg.de}}
\affiliation{Max Planck Institute for Solar System Research,
Katlenburg-Lindau, Germany.}
\author{A. Benslama}
\affiliation{Physics Department, Faculty of Science, Mentouri
University, Constantine, Algeria.}
\begin{abstract}
In the presence of a strong uniform magnetic field, we study the
influence of space noncommutativity on the electromagnetic waves
propagating through a quasi-static homogeneous plasma. In this
treatment, we have adopted a physical model which considers plasma
as quasi-neutral single fluid. By using noncommutative Maxwell
theory, the ideal magnetohydrodynamics (MHD) equations are
established, in which new equilibrium conditions are extracted. As
an empirical study, some attractive features of MHD waves behavior
are investigated. Furthermore, it is shown that the presence of
space noncommutativity enhances slightly the phase velocity of the
incompressive shear Alfv\'{e}n waves. In a compressible plasma, the
noncommutativity plays the role of an additional compression on the
medium, in which its relevant effect on the fast mode occurs for
highly oblique branchs, while the low effect appears when the
propagations are nearly parallel or anti-parallel. In addition, it
turned out that the influence of space deformation on the slow modes
is $\sim 10^{3}$ times smaller than that on the fast modes. The
space noncommutativity effect on the slow waves is negligible in low
plasma $\beta $ value, and could appear when $\beta $ is higher than
$0.1,$ thus the extreme modification occurs for oblique slow waves
propagating with angles between $30^{\circ }$ and $60^{\circ }$.
Finally, we comment on the possible effect of such waves on CMB
spectrum in photon-baryon plasma.
\end{abstract}




\maketitle

\section{Introduction:}

It is well known that magnetohydrodynamics (MHD) waves $\left[
1-3\right] $ play an important role in the field of space plasma. In
astrophysics, it remained for a long time the preferred theory in
the description of the dynamics of various astrophysical plasma
systems such as the formation of the solar corona which is
associated with the problem of the plasma heating and the solar wind
acceleration $\left[ 4-6\right] $. Indeed, many examples could be
given in the application of the MHD theory namely in space plasma,
like as the study of the linear properties of the fast magnetosonic
propagating in inhomogeneous plasma which is done by several authors
$\left[ 7-9\right] $ in order to model these waves in coronal loop.
T. K. Suzuki et al. $\left[ 10\right] $ proposed a collisionless
plasma in which the damped fast MHD waves are responsible of the
heating and acceleration of winds from rotating stars due to the
observational evidence for locally strong magnetic fields in stellar
atmospheres.

In cosmology, after the observation of the cosmic microwaves
background (CMB) radiations in 1965, it is believed that MHD played
a major role in shaping the radiation spectrum during the so-called
plasma epoch. In fact, the evidence that the plasma was magnetized
has been confirmed after the measurements of background magnetic
fields which are of the order of $\mu $Gauss. Moroever, most of the
theories predict that these magnetic fields are the amplified
remnants of a seed cosmological magnetic field generated in the
early Universe [11]. The presence of such field in the primordial
plasma influences the acoustic waves pattern of the CMB anisotropy
power spectrum $[12-14]$, e.g. Adams et al. $[13]$ argued that the
primordial density fluctuations that are generated in inflationary
Universe enter the horizon before the last photon scattering, and
initiate magneto-acoustic oscillations in the photon-baryon plasma
due to the presence of primordial magnetic fields. These
oscillations distort the primordial spectrum of fluctuations and
affect CMB anisotropy. Therefore, dealing with the features of MHD
wave dynamics is a significant part of cosmological plasma, and
because of the magnetic forces, the theory of MHD is more
complicated and fascinating than hydrodynamics itself due to the
influence of the magnetic field on the traditional sound waves which
consequently transform to slow and fast magneto-acoustic waves.
Furthermore, a new kind of wave appears, called Alfv\'{e}n wave,
which arises from magnetic tension and propagates along the field
lines without disturbing the thermal pressure or density of the
plasma.

On the other hand, there has been a large interest in the study of
physical phenomena on a non commutative (NC) spacetime. The idea of
\ spacetime noncommutativity is not new and was first discussed by
Snyder in 1947 $[15].$ At this time the theory of renormalization
was not yet well established and the goal was to introduce a natural
cut-off to deal with infinities in quantum field theory. However
this theory was plagued with several problems such as \ the
violation of unitarity and causality which make people abandoning
it. The appearance of such theory, baptized noncommutative geometry,
as a limit of string theory has generated a revival of interest for
this theory $[16,17]$.

In the framework of noncommutative geometry the position vector
$x^{\mu }$ is promoted to an operator $\hat{x}^{\mu }$ satisfying
the relation
\begin{equation}
\left[ \hat{x}^{\mu },\hat{x}^{\nu }\right] =i\theta ^{\mu \nu },
\end{equation}

where $\theta ^{\mu \nu }$ is a real, antisymmetric constant matrix
which has the dimension of area with elements of order ($\Lambda
_{NC})^{-2}$ \ in system unit $(\hbar =c=1)$. $\Lambda _{NC}$ is the
energy scale where the effects on the noncommutativity of spacetime
will be relevant.

The role of $\theta ^{\mu \nu }$ can be compared to that of the
Planck constant $\hbar $ which quantifies in quantum mechanics the
level of noncommutativity between space and momentum. In Moyal
algebra, the product of two arbitrary fields is defined by
$\ast $ product (the star or Moyal product) $[18,19]:$%
\begin{equation}
(f\ast g)(x)=\left[ \exp (\frac{i}{2}\theta ^{\mu \nu }\partial
_{x_{\mu }}\partial _{y_{\nu }})f(x)g(y)\right] _{x=y}.
\end{equation}

\bigskip Some cosmological effects of noncommutativity have been
studied. When spacetime is noncommutative on short distance scales,
this may have an imprint on early Universe physics, and leads to an
interesting consequence at a microscopic level. Indeed, it could be
one of possible scenarios that may cause the generation of the
density perturbations and primordial magnetic fields in the
inflationary Universe $[20]$. Such studies allow to predict some
bounds on noncommutativity scale $\Lambda _{NC}$ which may have a
temperature dependence [21].

\bigskip Our work is devoted to study the MHD waves by taking into account
the space modification, and focusing on the description of a
homogeneous plasma in a non-smooth space. The aim of this work is to
make a theoretical background in the field of NCMHD waves and seek
some future works to study this topic. Opening this new window
inevitably leads to deal with the interactions of space deformation
and plasma waves, which could be considered as a new experimental
area for testing space noncommutativity contribution.

\bigskip This paper is organized as follows. In the second section
we start with a brief review of noncommutative classical
electrodynamics, from which we derive the NCMaxwell equations. Then
in the third section, and by assuming a small $\theta $ matrix, we
establish the modified MHD equations to first $\theta -$parameter
for a single conductor medium. As an application, in the fourth
section we deal with a particularly interesting case of a high
conductor $\left( \sigma \rightarrow \infty \right) $ fluid which is
considered as plasma medium. New equilibrium conditions are deduced
as well as the main attractive features related to the NCMHD waves
are studied for a homogeneous plasma around an equilibrium state.
Finally,\ the obtained results are discussed and compared with those
known in usual space.

\section{Noncommutative Maxwell equations:}

It is understood that NC gauge field theories cause the violation of
Lorentz invariance when $\theta $ is considered as a constant
matrix, except if this matrix is promoted to a tensor related to the
contracted Snyder's Lie algebra $[22] .$ The problem of unitarity
appears also with time-space noncommutativities ($\theta ^{0i}\neq
0$) $\left[ 9,10\right] $. In particular, NC Maxwell theory loses
the causality due to the appearance of derivative couplings in the
Lagrangian with the Lorentz invariance exhibited by plane wave
solutions $[23].$

The free Maxwell action on noncommutative space is given by
\begin{equation}
S=-\frac{1}{4}\int dx\hat{F}_{\mu \nu }\ast \hat{F}^{\mu \nu }
\end{equation}%
where $\hat{F}_{\mu \nu }$ is the noncommutative strength field
\begin{equation}
\hat{F}_{\mu \nu }=\partial _{\mu }\hat{A}_{\nu }-\partial _{\nu }\hat{A}%
_{\mu }-ie\left[ \hat{A}_{\mu },\hat{A}_{\nu }\right] _{\ast },
\end{equation}%
where $e$ is the electric charge, and $\left[ \hat{A}_{\mu },\hat{A}_{\nu }%
\right] _{\ast }$ is the Moyal bracket defined as
\begin{equation*}
\left[ \hat{A}_{\mu },\hat{A}_{\nu }\right] _{\ast }=\hat{A}_{\mu }\ast \hat{%
A}_{\nu }-\hat{A}_{\nu }\ast \hat{A}_{\mu }.
\end{equation*}

According to the Seiberg-Witten map to the first $\theta $ order of
the NC
gauge and strength fields $[17] $ , we get%
\begin{eqnarray}
\hat{A}_{\mu } &=&A_{\mu }-\frac{e}{2}\theta ^{\alpha \beta
}A_{\alpha
}\left( \partial _{\beta }A_{\mu }+F_{\beta \mu }\right)  \notag \\
\hat{F}_{\mu \nu } &=&F_{\mu \nu }+e\theta ^{\alpha \beta }F_{\mu
\alpha }F_{\nu \beta }-e\theta ^{\alpha \beta }A_{\alpha }\partial
_{\beta }F_{\mu \nu },
\end{eqnarray}%
with $F_{\mu \nu }=\partial _{\mu }A_{\nu }-\partial _{\nu }A_{\mu
}$ is the usual strength electromagnetic field and $A_{\mu }$ is the
vector-potential.

Hence, from action $\left( 3\right) $ the Lagrangian in
four-dimensional
spacetime is%
\begin{eqnarray}
\mathfrak{L} &\mathit{=}&-\frac{1}{4}F_{\mu \nu
}^{2}+\frac{e}{8}\theta
^{\alpha \beta }F_{\alpha \beta }F_{\mu \nu }^{2}  \notag \\
&&-\frac{e}{2}\theta ^{\alpha \beta }F_{\mu \alpha }F_{\nu \beta
}F^{\mu \nu }+\mathit{O}\left( \theta
{{}^2}%
\right) +A_{\mu }J^{\mu },
\end{eqnarray}%
where we have added the external vector-current $J^{\mu }$ $=4\pi
(\rho _{q},\mathbf{j})$ and taken into consideration the integration
of the term $A_{\mu }\ast J^{\mu }$ over the whole spacetime which leads to $%
A_{\mu }J^{\mu }$ due to the integral property of star product
$[18]$.

By using the expressions of the electric field $E^{i}=F^{0i}$ and
the magnetic induction field $B_{k}=\frac{1}{2}\epsilon
_{ijk}F^{ij}$ $(\epsilon ^{123}=1),$ and considering $e\theta
^{ij}=\epsilon ^{ijk}\theta _{k}$ \ with $\theta ^{0i}=0$
(space-space noncommutativity), we can extract a non-linear
equation, one of the most important NC Maxwell equations in Gaussian
units as follows $[24] $
\begin{eqnarray}
\text{\ }\mathbf{\nabla .}\mathcal{E} &=&4\pi \rho _{q} \\
\text{ }\frac{\partial }{\partial t}\mathcal{E}\mathbf{-\mathbf{\nabla }%
\wedge }\mathcal{H} &=&-4\pi \mathbf{j},
\end{eqnarray}
where $\mathbf{j}$ is the current and $\rho _{q}$ is the charge
density. Eq. $\left( 8\right) $ represents the modified Ampere's
law. The displacement $\mathcal{E}$ and magnetic $\mathcal{H}$
fields are also given
by%
\begin{eqnarray}
\mathcal{E} &=&\mathbf{E+d},  \notag \\
\mathcal{H} &\mathbf{=}&\mathbf{B+h,}  \notag
\end{eqnarray}%
with%
\begin{eqnarray}
\mathbf{d} &=&\left( \mathbf{\theta }.\mathbf{B}\right)
\mathbf{E-}\left(
\mathbf{\theta }.\mathbf{E}\right) \mathbf{B-}\left( \mathbf{E}.\mathbf{B}%
\right) \mathbf{\theta }  \notag \\
\mathbf{h} &\mathbf{=}&\left( \mathbf{\theta }.\mathbf{B}\right) \mathbf{B+}%
\left( \mathbf{\theta }.\mathbf{E}\right)
\mathbf{E-}\frac{1}{2}\left(
\mathbf{E%
{{}^2}%
}-\mathbf{B%
{{}^2}%
}\right) \mathbf{\theta .}
\end{eqnarray}

The dual tensor $\tilde{F}_{\mu \nu }=\frac{1}{2}\epsilon _{\mu \nu
\alpha \beta }F^{\alpha \beta }$ ($\epsilon _{\mu \nu \alpha \beta
}$ is an antisymmetric tensor Levi-Civita) always satisfies the
equation $\partial
_{\mu }\tilde{F}^{\mu \nu }=0$ which implies that%
\begin{eqnarray}
\frac{\partial }{\partial t}\mathbf{B+\mathbf{\nabla }\wedge E} &=&0 \\
\text{\ }\mathbf{\nabla .B} &=&0,
\end{eqnarray}%
where the symbol $\mathbf{\wedge }$ denotes the vector product.

\bigskip The choice of the matrix $\theta ^{\mu \nu }$means that we are
dealing with a preferred frame in which the background
electromagnetic field related to $\theta $-matrix is only reduced to
a constant background magnetic field. \ As it was shown in several
works $\ [25-27]$, $\theta $ space-space noncommutativity preserves
the unitarity and is compatible with most works done on NC theories.

It turned out from the paper of Kruglov $[24]$ that the
electromagnetic waves solutions of the linear equations of the
classical electrodynamics are the solutions of the nonlinear wave
propagation equations of the electromagnetic fields derived from NC
Maxwell theory at $\mathbf{j}=0=\rho . $ Also, more features of the
classical waves propagating have been discussed by Z. Guralnik et
al. $[28]$. The authors deduced that the phase speed of these waves
is different from $c$ (with small modification) in case of a
transverse propagation with respect to the background magnetic field
induction, while the parallel propagation propagation is unchanged.
Furthermore, Y. Abe et al. $[29] $ studied a more general case of
the electric-magnetic duality symmetry within noncommutative Maxwell
theory, \ in which the polarizations of the propagating waves have
been discussed. T. Mariz et. al $[30] $ gave a detailed study on the
dispersion relation for plane waves in the presence of a constant
background electromagnetic field, the authors did not find any
restriction on the plane waves solution in the Seiberg-Witten
approach of noncommutative gauge theory which is not the case in
strictly Moyal approach, where they deduced that no plane waves are
allowed when time is noncommutative.

In our study, we focus on the classical behavior of the
electromagnetic waves propagating through a plasma with high
conductivity in the presence of both, a magnetic field, and NC
space. This treatment is based on the classical MHD theory which is
worked out in the framework of NC Maxwell theory.

\section{Noncommutative MHD equations:}

Let us start with the continuity equation that describes the flow
motion
\begin{equation}
\frac{\partial \rho }{\partial t}+\mathbf{\nabla }.\left( \rho \mathbf{V}%
\right) =0
\end{equation}%
with $\rho $ is the mass density of the medium, $\mathbf{V}$ its
velocity. If this fluid has the ability to carry a density current
$\mathbf{J}$ (conductor), then the rising Ampere force is
$\mathbf{J\times B,}$ once the magnetic field $\mathbf{B}$ is
present, and the plasma is a quasi-neutral in large scale greater
than Debye length, consequently the electric volume force $\rho
_{q}\mathbf{E}$ vanishes, and
the moment fluid equation becomes%
\begin{equation}
\rho \left( \frac{\partial \mathbf{V}}{\partial t}\mathbf{+}\left( \mathbf{V}%
.\nabla \right) \mathbf{V}\right) =-\nabla p+\mathbf{j\wedge B},
\end{equation}%
with $p$ is the pressure which acts on the boundaries of the
infinitesimal fluid volume.

\bigskip In case of a medium with high conductivity $\left( \sigma
\rightarrow \infty \right) $ (ideal plasma), the known Ohm's law for
a moving conductor which describes the current $\mathbf{J}$ in terms
of magnetic and electric fields is reduced to the following simple
relationship
between these fields%
\begin{equation}
(\mathbf{E+V\wedge B})=0\mathbf{.}
\end{equation}

Since the quasi-neutrality is assumed, the equation $\nabla .\mathcal{E}%
\approx 0$ does not constitute a dynamical evolution equation. Also
the equation $\nabla .\mathbf{B}=0$ is only a constraint, not an
evolution equation since it does not include time derivative. In
case of non-relativistic MHD approximation where the conducting
fluid moves very slowly, this term is absent$\left( \frac{\partial
\mathbf{E}}{\partial t}\sim 0\right) $ which is the limit of slow
motion and large scale spatial derivatives, the displacement current
is always negligible, and from Eq. $\left( 8\right) ,$ we get
\begin{equation}
\mathbf{j=}\frac{1}{4\pi }\nabla \mathbf{\wedge }\mathcal{H}\mathbf{\mathbf{-%
}}\frac{1}{4\pi }\frac{\partial \mathbf{d}}{\partial t}.
\end{equation}%
\bigskip In order to express the relations of \ $\mathbf{d}$ and
$\mathbf{h}$ given in Eq. $\left( 9\right) $ in terms of
$\mathbf{\theta }$, $\mathbf{B}$ and $\mathbf{V}$, we use Eq. $(14)$
to extract the following relationships
up to the first order $\theta $%
\begin{eqnarray}
\mathbf{d} &=&\left( \mathbf{-}\left( \mathbf{\theta
}.\mathbf{B}\right)
\left( \mathbf{V\wedge B}\right) \mathbf{+}\left( \left( \mathbf{V\wedge B}%
\right) .\mathbf{\theta }\right) \mathbf{B}\right)  \notag \\
\mathbf{h} &\mathbf{=}&\mathbf{\left( \mathbf{\theta }.\mathbf{B}\right) B+}%
\left( \left( \left( \mathbf{V\wedge B}\right) .\mathbf{\theta
}\right)
\left( \mathbf{V\wedge B}\right) \right.  \notag \\
&&\left. \mathbf{-}\frac{1}{2}\left( \left( \mathbf{V\wedge B}\right) ^{2}-%
\mathbf{B%
{{}^2}%
}\right) \mathbf{\theta }\right) .
\end{eqnarray}

To establish the NCMHD equations, it is useful to express the current $%
\mathbf{j}$ as functions of $\mathbf{\mathbf{B}}$ and $\mathbf{V.}$
By using
Eqs. $\left( 14\right) $ and $\left( 10\right) \mathbf{,}$ we then obtain%
\begin{equation}
\frac{\partial \mathbf{B}}{\partial t}\mathbf{=}\nabla \mathbf{\mathbf{%
\wedge }}\left( \mathbf{V\wedge B}\right).
\end{equation}

\bigskip When $\mathbf{\theta }$ goes to zero, Eq. $\left(
13\right) $ tends to the usual known momentum equation.

\bigskip Based on the NCMHD equations, we study the electromagnetic waves
behavior using a linear mode analysis around the equilibrium. First,
we should determine the equilibrium condition of the considered
ideal plasma in the presence of noncommutativity. Plasma is said to
be in the equilibrium if $\mathbf{V}=0$ and none of the variables
depend on time, so from Eq. $\left(
13\right) ,$ we get%
\begin{equation}
\nabla p=\mathbf{j}\wedge \mathbf{B.}
\end{equation}%
Using the relationships $\left( 15\right) $ and $\left( 16\right) $,
we
deduce that%
\begin{eqnarray}
\nabla p &=&-\nabla \left( \frac{\left( \frac{1}{3}\mathbf{\left( \mathbf{%
\theta }.\mathbf{B}\right) +}1\right) B^{2}}{4\pi }\right)  \notag \\
&&+\frac{1}{4\pi }B(1+\mathbf{\left( \mathbf{\theta }.\mathbf{B}\right) })%
\mathbf{\nabla }B+\frac{B^{2}}{4\pi }\mathbf{\nabla \left( \mathbf{\theta }.%
\mathbf{B}\right) }  \notag \\
&&\frac{1}{4\pi }\left( \mathbf{B.\nabla }\right) \left(
\mathbf{B+\left(
\mathbf{\theta }.\mathbf{B}\right) B+}\frac{\mathbf{B%
{{}^2}%
}}{2}\mathbf{\theta }\right) ,
\end{eqnarray}%
where the first term represents the modified magnetic pressure and
the rest terms are the modified magnetic tension. In the equilibrium
conditions, the evolution equation of the magnetic field in the
noncommutative space is different from that given in usual space.
Certainly, the homogeneous case of constant $\mathbf{B}$ and $p$ are
one of the possible solutions of Eqs. $\left( 19\right) .$ In the
next treatment of the electromagnetic waves propagation through a
plasma around the equilibrium, we will assume that the component of
the background
field along the direction of the magnetic field, which means that $\mathbf{%
\theta }$ is parallel to $\mathbf{B}$.

\section{Noncommutative MHD waves}

It is well known that the waves play an important role in the
propagative phenomena related to the plasma physics. The following
treatment is mainly devoted to the description of the wave
properties of plasma in noncommutative space within MHD
approximation by a linear mode analysis. Notice that, the equations of ideal NCMHD $\left( 12\right) ,$ $%
\left( 13\right) $ and $\left( 17\right) $ are highly nonlinear and
self-consistent.

Let us study an homogeneous plasma (in equilibrium state) under a
constant magnetic field $\mathbf{B}_{0}$, in which all the
parameters do not depend on coordinates, i.e. $\rho _{0}$ and $p_{0}$ are constant with $%
\mathbf{V}_{0}=0$. Hence, all quantities $Q\left( \rho
,p,\mathbf{V}\text{ and }\mathbf{B}\right) $ can thus be written as
a sum of an equilibrium term and a small first-order perturbation%
\begin{equation}
Q\left( r,t\right) =Q_{0}\left( r\right) +\epsilon Q_{1}\left(
r,t\right) .
\end{equation}%
Higher order terms describing the perturbations are neglected. We
linearize the NCMHD equations by inserting the Fourier
transformation of all quantities
\begin{equation}
Q_{1}\left( r,t\right) =\int dk\tilde{Q}_{1}\exp i\left(
\mathbf{k.r-}\omega t\right) \text{ }
\end{equation}%
into eqs. $\left( 12\right) ,$ $\left( 13\right) $ and $\left(
17\right) $. Then, we extract the terms of first order $\epsilon $
with neglecting all remaining terms of order $\epsilon
{{}^2}%
$ and higher as follows
\begin{eqnarray}
\tilde{\rho}_{1} &=&\frac{\rho _{0}}{\omega }\mathbf{k.\tilde{V}}_{1}, \\
\omega \rho _{0}\mathbf{\tilde{V}}_{1} &=&\frac{v_{s}^{2}\rho _{0}}{\omega }%
\mathbf{k}\left( \mathbf{k.\tilde{V}}_{1}\right) -\frac{e^{\theta B_{0}}}{%
4\pi }\mathbf{C}\wedge \mathbf{B}_{0}  \notag \\
&&-\frac{\left( \mathbf{B}_{0}.\mathbf{\tilde{B}}_{1}\right) }{4\pi
}\left(
\mathbf{k\wedge \theta }\right) \wedge \mathbf{B}_{0}, \\
\omega \mathbf{C} &=&-\left( \mathbf{k\wedge \tilde{V}}_{1}\right)
\left(
\mathbf{k.B}_{0}\right) +\left( \mathbf{k\wedge B}_{0}\right) \left( \mathbf{%
k.\tilde{V}}_{1}\right)
\end{eqnarray}%
with $1+\theta B_{0}\approx e^{\theta B_{0}}$ and $\mathbf{C=k\wedge \tilde{B%
}}_{1}$.

Notice that we have considered the static magnetic field
$\mathbf{B}_{0}$ is parallel to $\mathbf{\theta }$ which is
proportional to the background magnetic field$.$ A polytropic
adiabatic law which gives the relationship between the pressure $p$
and the density $\rho $ has been taken into account
\begin{equation}
p=p_{0}\left( \frac{\rho }{\rho _{0}}\right) ^{\gamma }.
\end{equation}

As a consequence, $p_{1}=v_{s}^{2}\rho _{1}$, from which, we derive
the expression of the square of sound speed $v_{s}^{2}=\frac{\gamma
p_{0}}{\rho _{0}},$ with $\gamma $ is the adiabaticity index.

The equations $\left( 22\right) ,\left( 23\right) $ and $\left(
24\right) $ are a homogeneous set of 6 independent equations for 6
variables since on the other hand, we have the constraint
$\mathbf{k.\tilde{B}}_{1}=0$ which indicates a transverse
propagation of electromagnetic waves$.$ From Eq. $\left( 22\right)
$, we deduce that the density variations are only related to the
velocity components along the wave vector, and similarly to the
commutative case, the study of the waves in an incompressible plasma
$\left( \rho _{1}=p_{1}=0\right) $ shows that the velocity motion is
transversal with respect to the wave vector $\mathbf{k}$ and
consequently, the fluctuating
magnetic field $\mathbf{\tilde{B}}_{1}$ is perpendicular to $\mathbf{\mathbf{%
B}_{0}.}$

In order to analyze the various modes arising from the perturbed
homogeneous plasma, it is worth focusing on the dispersion relations
which connect the
wave phase velocity $\mathbf{v}_{ph}=\frac{\omega }{k%
{{}^2}%
}\mathbf{k}$ with the measured quantities related to the plasma
features themselves. By substituting the vector product of eq.
$\left( 23\right) $
with $\mathbf{k}$ into Eq. $\left( 24\right) ,$ we get%
\begin{eqnarray}
&&\left( 1-\frac{e^{\theta B_{0}}\left( \mathbf{k.B}_{0}\right)
{{}^2}%
}{4\pi \rho _{0}\omega ^{2}}\right) \mathbf{C-}\frac{1}{\omega
}\left( \mathbf{k\wedge B}_{0}\right) \left(
\mathbf{k.\tilde{V}}_{1}\right)  \notag
\\
&=&\frac{e\left( \mathbf{k.B}_{0}\right)
{{}^2}%
}{4\pi \rho _{0}\omega ^{2}}\left( \mathbf{B}_{0}.\mathbf{\tilde{B}}%
_{1}\right) \left( \mathbf{k\wedge \theta }\right) \text{ }
\end{eqnarray}

At this stage, it is worth analyzing the Alfv\'{e}n and magnetosonic
waves propagating in incompressible and compressible plasmas, and
compare them with those obtained in commutative space. Let us
separately treat the two different states of plasma when the
propagation is not purely sonic $\left( v_{ph}\neq v_{s}\right) $.
Firstly, we consider the incompressible case of plasma,
i.e. $\mathbf{B}_{0}.\mathbf{B}_{1}=0$ which leads to%
\begin{equation*}
\mathbf{C\wedge }\left( \mathbf{k\wedge \mathbf{B}_{0}}\right) \neq
0.
\end{equation*}%
From the vector product of Eq. $\left( 26\right) $ with $\left( \mathbf{%
k\wedge \mathbf{B}_{0}}\right) $, we obtain the first dispersion
relation for an incompressible plasma

\begin{equation}
\left( 1-\frac{e^{\theta B_{0}}\left( \mathbf{k.B}_{0}\right)
{{}^2}%
}{4\pi \rho _{0}\omega ^{2}}\right) =0,
\end{equation}%
where we can get the norm of wave phase velocity in the presence of
noncommutativity%
\begin{equation}
v_{ph}^{2}=\left( v_{A}^{nc}\right) ^{2}\cos ^{2}\alpha \text{ }
\end{equation}%
with $\left( v_{A}^{nc}\right) ^{2}=e^{\theta B_{0}}v_{A}^{2}$ is
the modified Alfv\'{e}n velocity and $v_{A}^{2}=B_{0}^{2}/4\pi \rho
_{0}$ is the usual Alfv\'{e}n velocity. Indeed, in case of\ a
parallel propagation, the modified incompressible Alfv\'{e}n waves
can be deduced,
\begin{equation}
v_{ph}^{2}=\left( v_{A}^{nc}\right) ^{2}.
\end{equation}

Note that the Alfv\'{e}n mode undergoes a modification coming from
space deformation. This modification plays a role of another
velocity which is added to the usual Alfv\'{e}n velocity,
consequently, enhances the value of this latter. In fact, the
influence of the noncommutativity is considerable when we deal with
an incompressible plasma emerged in a strong magnetic field
$\mathbf{B}_{0}.$
\begin{figure}[tbp]
\includegraphics[width=8cm,height=70mm]{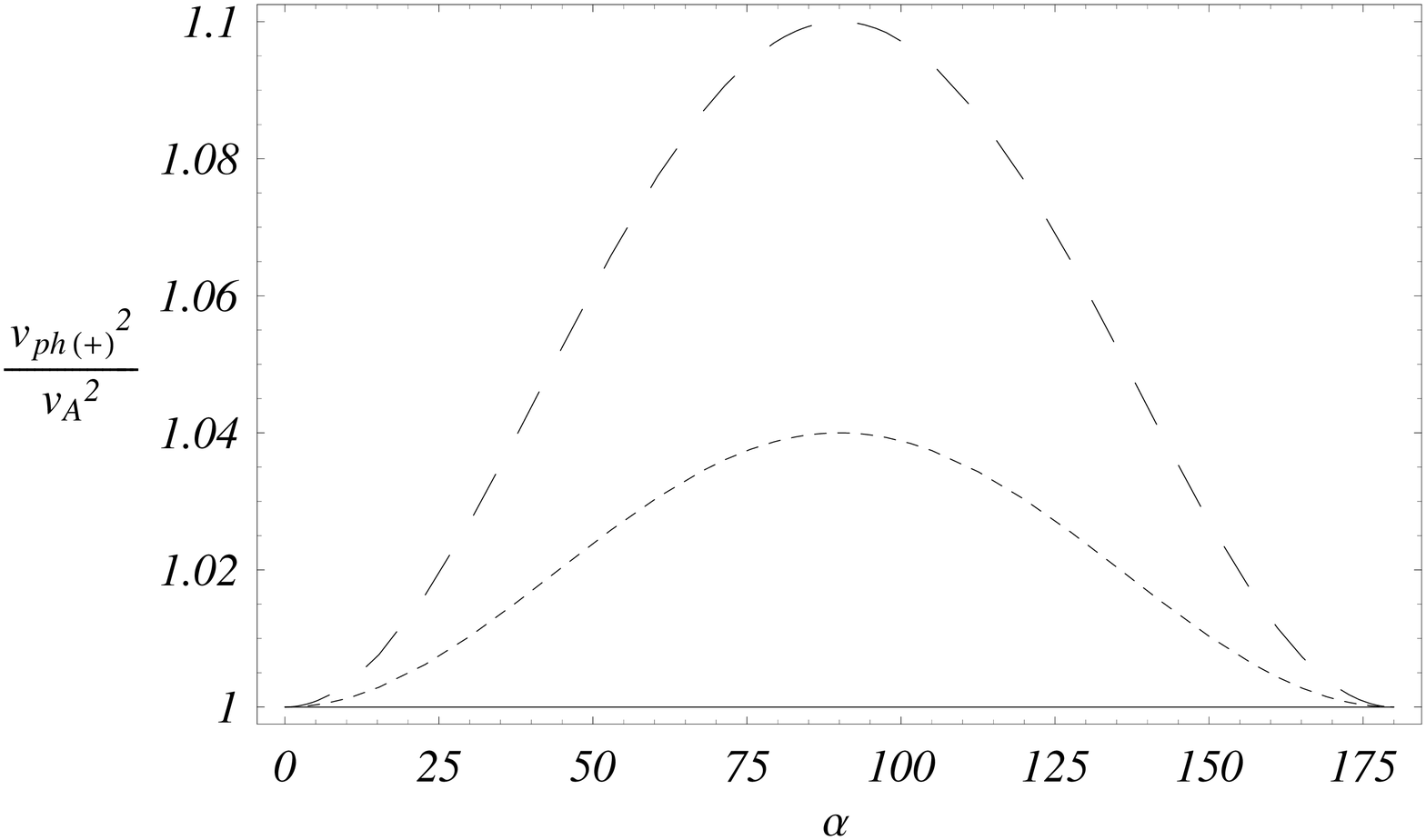}
\caption{The variation of the square phase speed of fast mode
normalized to
the square Alfven velocity as a function of the propagation angle $\protect%
\alpha $ in the absence of noncommutativity $(\protect\theta =0)$ for $%
\protect\beta =0.1$ (dashed line), $\protect\beta =0.04$ (dotted line) and $%
\protect\beta =0$ (line). }
\end{figure}
\begin{figure}[tbp]
\includegraphics[width=8cm,height=70mm]{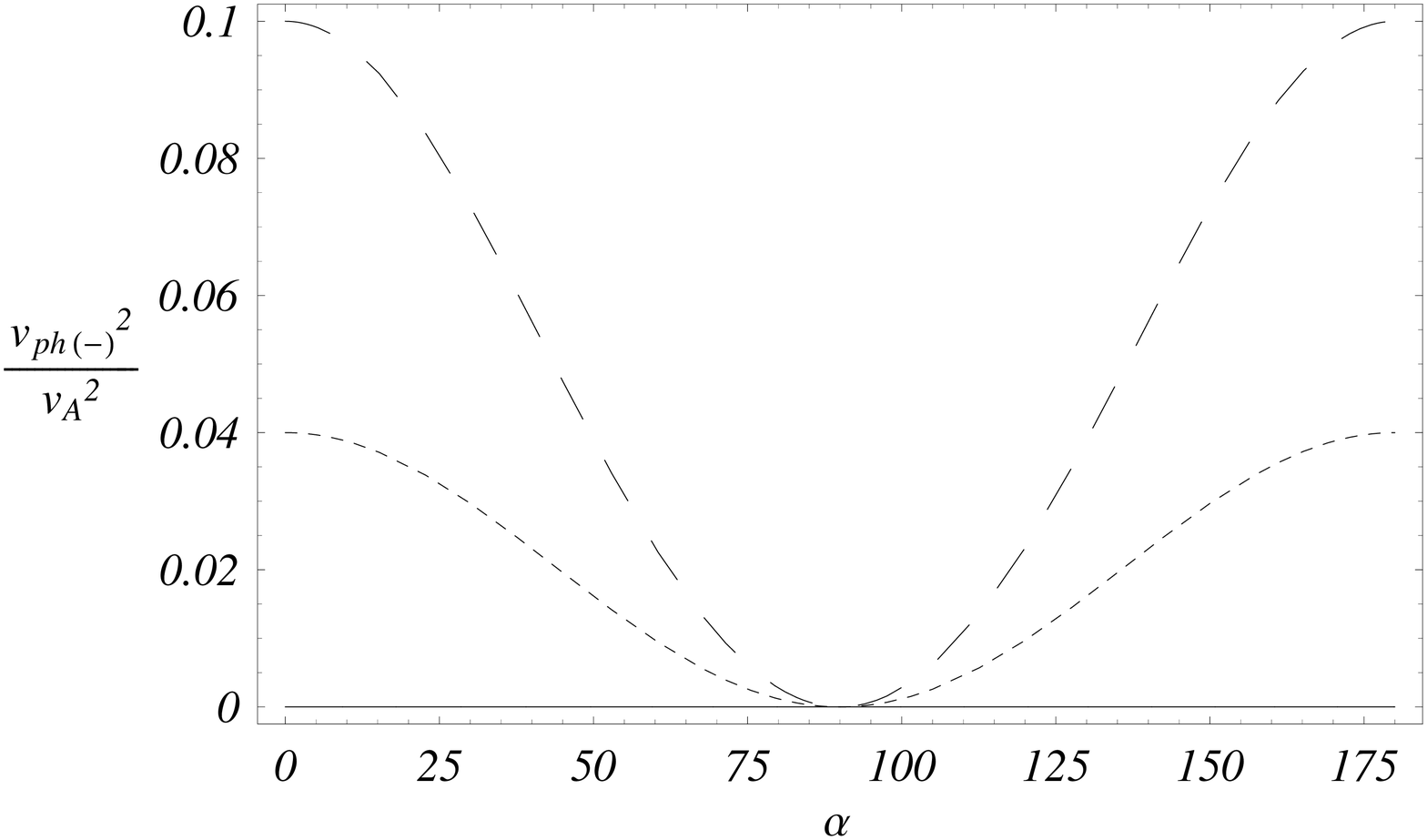}
\caption{ The variation of the square phase speed of slow mode
normalized to
the square Alfven velocity as a function of the propagation angle $\protect%
\alpha $ in the absence of noncommutativity $(\protect\theta =0)$ for $%
\protect\beta =0.1$ (dashed line), $\protect\beta =0.04$ (dotted line) and $%
\protect\beta =0$ (line). }
\end{figure}

Secondly, if the plasma is compressible, this means that $\mathbf{B}_{0}.%
\mathbf{B}_{1}\neq 0,$ therefore, by performing the scalar product of Eq. $%
\left( 26\right) $ with $\left( \mathbf{k\wedge
\mathbf{B}_{0}}\right) $, we obtain the second dispersion relation
for a compressible plasma
\begin{eqnarray}
&&\left( 1-\frac{e^{\theta B_{0}}\left( \mathbf{k.B}_{0}\right)
{{}^2}%
}{4\pi \rho _{0}\omega ^{2}}\right) \left( 1-\frac{k%
{{}^2}%
v_{s}%
{{}^2}%
}{\omega
{{}^2}%
}\right) =  \notag \\
&&\frac{1}{4\pi \rho _{0}\omega
{{}^2}%
}\left( \mathbf{k\wedge \mathbf{B}_{0}}\right) ^{2}\left[ e^{\theta B_{0}}+%
\frac{2}{k%
{{}^2}%
}\left( \mathbf{k\wedge \theta }\right) .\left( \mathbf{k\wedge B}%
_{0}\right) \right]  \notag \\
&&+\frac{\left( \mathbf{k.B}_{0}\right)
{{}^2}%
}{4\pi \rho _{0}\omega ^{2}k%
{{}^2}%
}\left[ \left( \mathbf{k}\wedge \mathbf{\theta }\right) .\left( \mathbf{k}%
\wedge \mathbf{\mathbf{B}_{0}}\right) \right] \left( 1-\frac{k%
{{}^2}%
v_{s}%
{{}^2}%
}{\omega
{{}^2}%
}\right) ,
\end{eqnarray}

from which, we extract the following fast and the slow modes%
\begin{eqnarray}
\left( v^{nc}\right) _{ph\left( \pm \right) }^{2}
&=&\frac{1}{2}\left[ \pm \left( \left( \left( v_{A}^{nc}\right)
^{2}\exp \left( \theta B_{0}\sin
^{2}\alpha \right) +v_{s}^{2}\right) ^{2}\right. \right.  \notag \\
&&\left. -4\left( v_{A}^{nc}\right) ^{2}v_{s}^{2}\exp \left( \theta
B_{0}\sin ^{2}\alpha \right) \cos
{{}^2}%
\alpha \right) ^{1/2}  \notag \\
&&\left. +\left( \left( v_{A}^{nc}\right) ^{2}\exp \left( \theta
B_{0}\sin ^{2}\alpha \right) +v_{s}^{2}\right) \right].
\end{eqnarray}

The signs $(+)$ and $(-)$ in eq. $\left( 31\right) $\ indicate the
fast and slow magnetosonic (or magneto-acoustic) waves which arise
from the coupling between magnetic compression (Alfv\'{e}nic) and
medium compression (sonic). We note that the influence of
noncommutativity on the magneto-acoustic waves involves again the
modified Alfv\'{e}n velocity $\left( 29\right) $.

Let us define the quantity $\Delta _{\pm }$
\begin{equation}
\Delta _{\pm }=\frac{\left( v^{nc}\right) _{ph\left( \pm \right)
}^{2}-v_{ph\left( \pm \right) }^{2}}{\theta B_{0}v_{A}^{2}},
\end{equation}%
that corresponds to the degeneracy rate of the fast and slow modes
$\left( v^{nc}\right) _{ph\left( \pm \right) }^{2}$ in the presence
of space noncommutativity from the usual modes $v_{ph\left( \pm
\right) }^{2}$ normalized to $\theta B_{0}$ quantity and the square
of the usual Alfv\'{e}n velocity $v_{A}^{2}$. Notice that
$v_{A}^{2}$ and $v_{ph\left( \pm \right) }^{2}$ are respectively
$\left( v_{A}^{nc}\right) ^{2}$ and $\left( v^{nc}\right) _{ph\left(
\pm \right) }^{2}$ in the absence of noncommutativity ($\theta =0).$

\bigskip \bigskip By considering a small value of $\theta B_{0}=10^{-5}$
which is a consequence of a strong value of the mean magnetic field,
we plot $\Delta _{+}$ and $\Delta _{-}$ respectively in fig. 3 and
fig. 4 in function of the propagation angle $\alpha $ for a
different values of plasma $\beta =\frac{v_{s}^{2}}{v_{A}^{2}}$.

It is well known that the strength of the value of the magnetic
field plays the major role in making the effect of the space
deformation more relevant on the waves in plasma medium when the
noncommutativity scale is relatively high.

\begin{figure}[tbp]
\includegraphics[width=8cm,height=70mm]{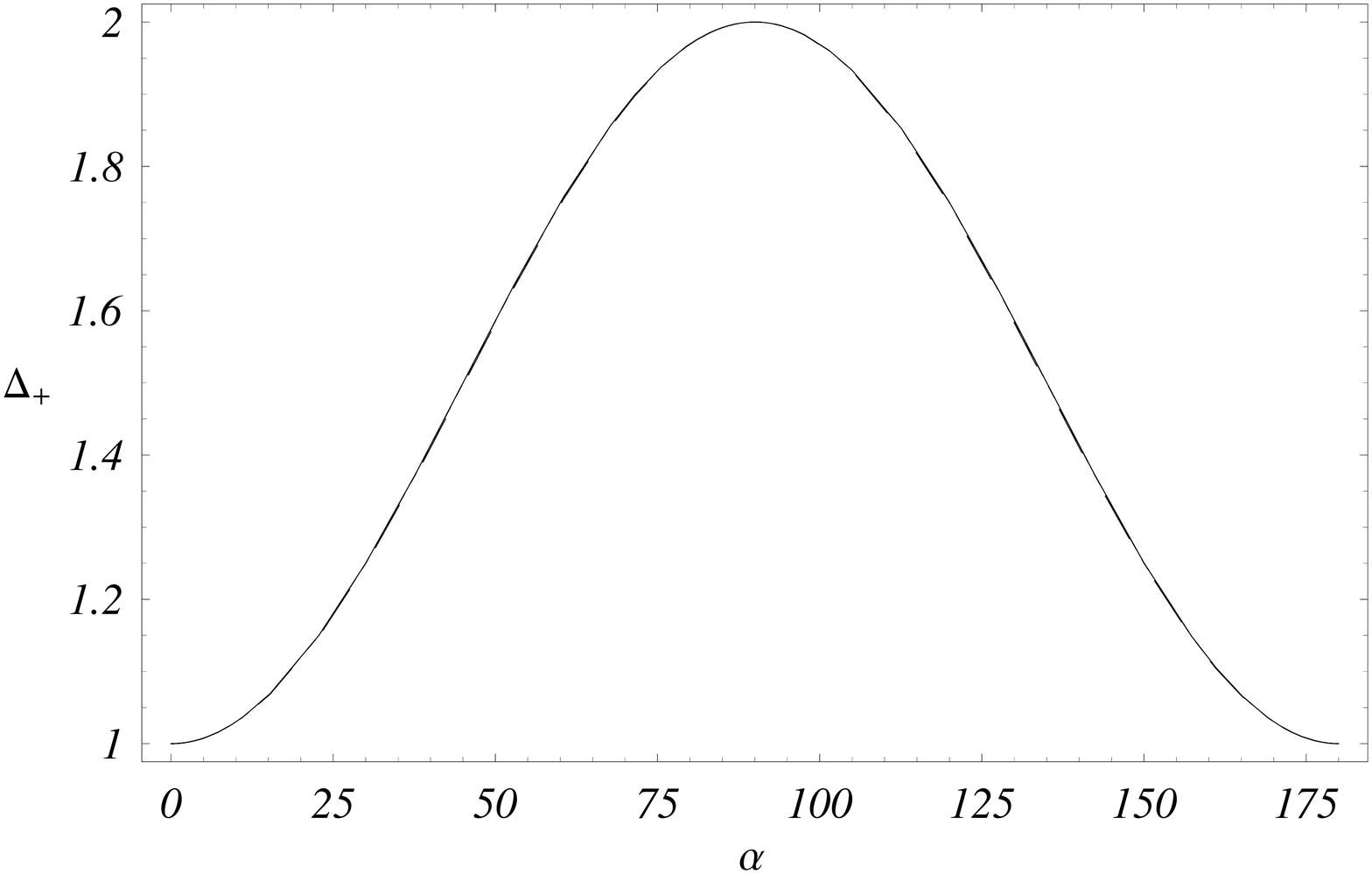}
\caption{The variation of the degeneracy rate $\Delta_{+}$ for
oblique fast
waves as a function of the propagation angle $\protect\alpha$ for $\protect%
\beta= 0.1, 0.04, 0$.}
\end{figure}
\begin{figure}[tbp]
\includegraphics[width=8cm,height=70mm]{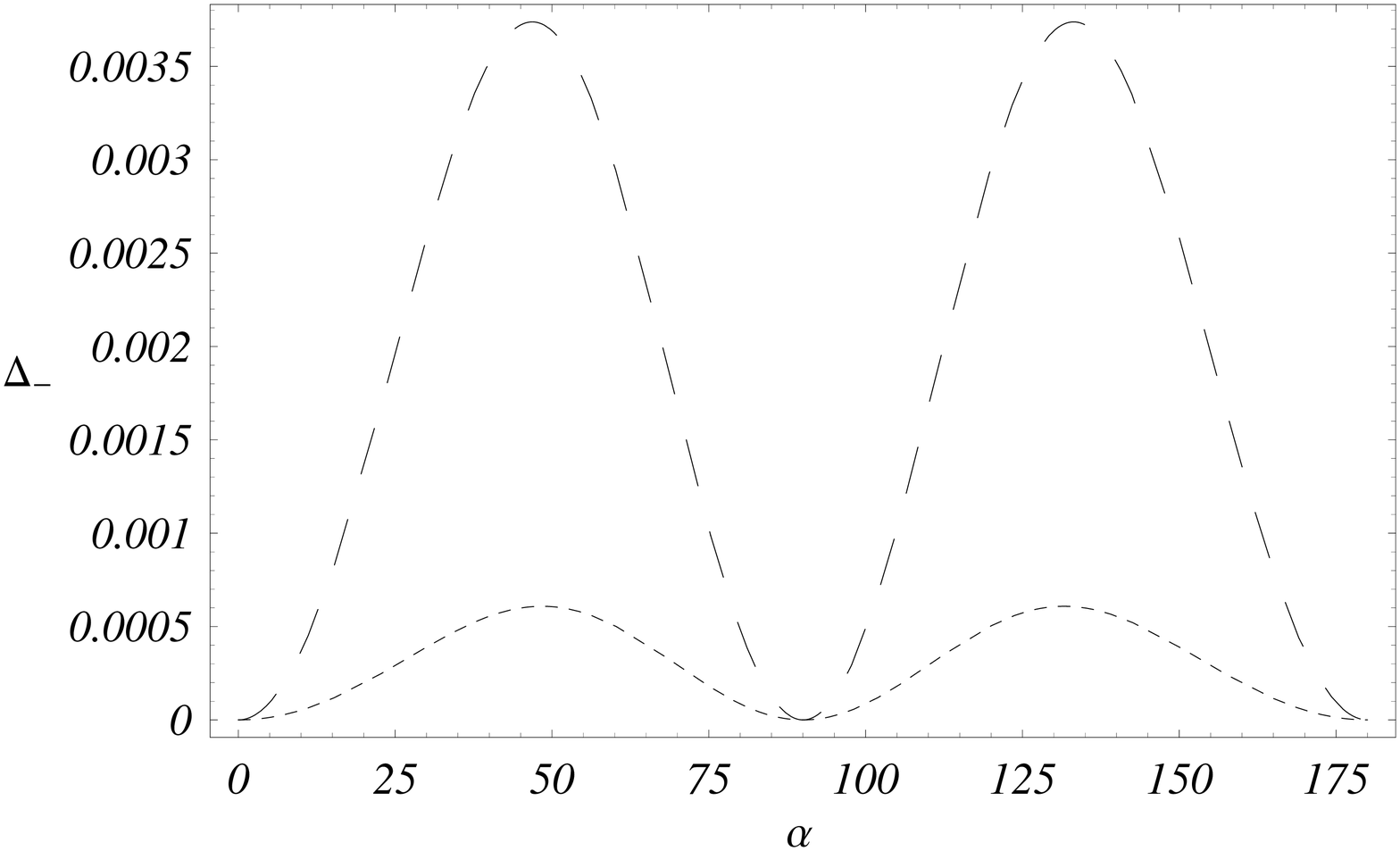}
\caption{The variation of the degeneracy rate $\Delta_{-}$ for
oblique slow
waves as a function of the propagation angle $\protect\alpha$ for $\protect%
\beta= 0.1$ (dashed line), $\protect\beta= 0.04$ (dotted line) and $\protect%
\beta=0 $ (line). }
\end{figure}

The plasma $\beta $ variation in fixed mean magnetic field $B_{0}$
becomes
proportional to the pressure $p_{0}$ by the adiabatic index $\gamma ,$ then $%
\beta $ increases when the plasma is initially strongly compressed.
However in our study, we consider that the magnetic pressure is
dominant in such way that $\beta <1$.

According to the fig. 3, it turns out \ that in compressible plasma,
the phase velocity related to the fast waves is slightly enhanced
due to space noncommutativity. This modification which is
represented by the rate $\Delta _{+}$, has a very slow dependence on
plasma $\beta $ variation. This extreme modification on the fast
which is around $\sim 2$ which occurs when the
propagation is perpendicular ($\alpha =90%
{{}^\circ}%
$) with respect to the mean magnetic field, and becomes smaller with
$\Delta _{+}\sim 1$ when the fast waves is nearly Alfv\'{e}nic in
parallel or anti-parallel. Also, the noncommutativity effect on the
fast mode is proportional to the phase speed of the wave which
varies as a function of the propagation angle. In contrast, from
fig. 4, the quantity $\Delta _{-}$ that corresponds to modification
rate of the slow mode has a high dependence on plasma $\beta $
variation. Although this rate is much smaller comparing with
$\Delta _{+}$, the noncommutative effect could slightly appear when plasma $%
\beta $ is higher than $0.04$, which means that the noncommutativity
affects the slow mode as long as the plasma is strongly compressed
(high $p_{0}$). The extreme
modification on slow mode occurs mainly in oblique propagations between $30%
{{}^\circ}%
$ and $60%
{{}^\circ}%
$, and as it is expected from fig.2, the noncommutativity effect
vanishes
for $\alpha =90%
{{}^\circ}%
$ at which originally there is no perpendicular propagation of the
slow mode. In addition, no modification appears on parallel and
anti-parallel slow waves which correspond to the maximum phase
velocity of this mode. This means that no proportionality between
noncommutativity effect and the phase velocity of the slow wave.

It is important to discuss the interesting physical case when $\beta
\rightarrow 0$ , which means a total domination of the magnetic
pressure on
compressible plasma due to the strength of the mean constant magnetic field $%
B_{0}$ or the plasma is not relatively enough compressed . In fact,
according to fig. 1, in the absence of noncommutativity ($\theta
=0)$, the fast mode converges to the usual Alfv\'{e}n mode
($v_{+}\approx v_{A}$) for any propagation angle, but when $\theta
\neq 0$, it is clear from fig.3 that the effect of space
noncommutativity does not change and is nearly the same one for
$\beta \neq 0.$ Therefore, this effect plays a role of medium
compression which rises the fast mode from the Alfv\'{e}nic one.
While in fig.2, as it is expected, the slow mode vanishes in case of
$\beta =0$ which is equivalent to the incompressible plasma case.

\section{Conclusion and discussion}

In this letter, we have studied the propagative phenomenon of the
electromagnetic waves in a homogeneous plasma under the effect of a
magnetic field and space-space uncertainty. By using a physical
model which considers plasma as a high conductor single fluid
medium, the ideal NCMHD equations are established with the help of
NC Maxwell theory. Because of a quasi-static motion of the fluid, we
neglected the relativistic effect which is very small and it is
below the order of $(\theta B_{0})^{2}$. In this treatment, the
behavior of the electromagnetic waves propagating in this medium are
studied in both cases, incompressible and compressible plasmas. It
turned out that in an incompressible plasma, the Alfv\'{e}n mode
undergoes a modification which appears as a small additional
velocity which enhances the value of the usual Alfv\'{e}n velocity
in commutative space.

On the other hand, it is deduced that the influence of space
noncommutativity on the fast waves in a compressible plasma is
proportional to the phase velocity. Moreover, the high oblique fast
waves undergo a strong modification which is about $2$ times higher
than the perturbed parameter $\theta B_{0}$, and a weak modification
occurs in case of nearly parallel and anti-parallel propagation. For
the slow mode, the influence of space noncommutativity is very small
especially when $\beta $ is low than $0.04$. This effect could
slightly appear when $\beta $ is beyond the value of $0.1$, at
which, the extreme modification on the slow mode occurs in oblique propagations between $30%
{{}^\circ}%
$ and $60%
{{}^\circ}%
$, and vanishes in the parallel one. In addition, for $\beta
\rightarrow 0$ there is no influence of space noncommutativity on
slow mode while this influence on the fast mode is nearly identical
to that for \ $\beta \neq 0$ which is higher in perpendicular
propagation.

 The impact of space noncommutativity on cosmological MHD waves
properties may lead to new consequences on the study of the
influence of such waves on CMB temperature spectrum. Although the
effect of the space noncommutativity may decrease after the
inflationary universe due to the decreasing in energy scale, the
presence of a primordial magnetic field excites any possible effect
of space-space uncertainty on the primordial photon-baryon plasma
before the last scattering. Any role of space noncommutativity at
that time depends on its scale $\Lambda_{NC}$. Several scenarios
based on space noncommutativity, aimed to explain the mechanism
behind the generation of the primordial magnetic field, hence,
possible constraints on $\Lambda_{NC}$ parameter have been involved.
In fact, the possibility that $\Lambda_{NC}$ has a temperature
dependence has been discussed in [20, 21], where the authors
mentioned that the world is commutative at low temperature but
becomes more noncommutative once the temperature is higher than a
certain threshold temperature $T_{0}$. In ref. [21], an intensive
discussion on the temperature dependence of $\theta$-parameter based
on the constraints on the primordial magnetic field which is
$B(T=10Mev)= 10^{-8} Gev^{2}$ at the beginning of nucleosynthesis
(see ref. [31]), the authors argued that the presence of
noncommutativity may be not so efficient beyond nucleosynthesis
scale. However, its effect cannot be omitted due to high temperature
especially when the radiations dominate the plasma. Hence, this
drives our attention, that this effect can be treated as a
perturbative correction in the study of the distortion of the
primordial spectrum of fluctuations by MHD waves before the last
scattering. As we have seen in this letter, the space
noncommutativity influences the compressible oscillations depending
on their propagation angle, on the other hand, the fact that the
velocity of the fast waves depends on the propagation angle between
the wave-number and the magnetic field, the CMB anisotropy would be
affected $[13]$. This could reduce the impact of noncommutativity on
some branches of the MHD waves, therefore, this leaves imprints on
the primordial spectrum of the fluctuations anisotropy. Furthermore,
such an influence may also correct the magnitude of the primordial
magnetic field, this is because the noncommutativity can play a role
of an additional magnetic field applied on the medium. In addition,
dealing with the influence of the NCMHD waves on CMB radiations may
provide a better estimation of the scale of $\theta$ parameter
during plasma epoch. More details will be given in our future works.

\end{document}